\documentclass[journal]{IEEEtran}
\ifCLASSINFOpdf
\else
\fi

\usepackage{xkeyval}  
\usepackage[textsize = footnotesize]{todonotes}
\usepackage{xcolor}

\usepackage[bookmarks=true,bookmarksnumbered=true,
bookmarksopen=false, breaklinks=false,pdfborder={0 0 0},
pdfpagemode=UseNone,backref=false,colorlinks=false] {hyperref}

\usepackage{mathtools}
\usepackage{cuted}

\usepackage{amsthm}

\usepackage{float}

\usepackage{textcomp} 
\usepackage{gensymb}
\usepackage{amsmath}
\usepackage{siunitx} 
\usepackage{listings}
\usepackage{mathtools}

\makeatletter
\def\resetMathstrut@{%
  \setbox\z@\hbox{%
    \mathchardef\@tempa\mathcode`\[\relax
    \def\@tempb##1"##2##3{\the\textfont"##3\char"}%
    \expandafter\@tempb\meaning\@tempa \relax
  }%
  \ht\Mathstrutbox@\ht\z@ \dp\Mathstrutbox@\dp\z@}
\makeatother
\begingroup
  \catcode`(\active \xdef({\left\string(}
  \catcode`)\active \xdef){\right\string)}
\endgroup
\mathcode`(="8000 \mathcode`)="8000

\usepackage{url}

\usepackage{multirow}

\usepackage{subcaption}

\usepackage{graphicx}
\usepackage{epstopdf}

\usepackage[]{algorithm2e}

\usepackage{xkeyval}  
\usepackage[textsize = footnotesize]{todonotes}
\presetkeys{todonotes}{inline}{}

\renewcommand{\theenumi}{\alph{enumi}}

\usepackage{cleveref}
\AtBeginDocument{\renewcommand{\ref}[1]{\Cref{#1}}}
\Crefformat{equation}{#2(#1)#3}
\Crefformat{figure}{#2Fig. #1#3}

\newcommand{\vect}[1]{\boldsymbol{\mathbf{#1}}}

\usepackage{scalerel}
\usepackage{tikz}
\usetikzlibrary{svg.path}

\definecolor{orcidlogocol}{HTML}{A6CE39}
\tikzset{
	orcidlogo/.pic={
		\fill[orcidlogocol] svg{M256,128c0,70.7-57.3,128-128,128C57.3,256,0,198.7,0,128C0,57.3,57.3,0,128,0C198.7,0,256,57.3,256,128z};
		\fill[white] svg{M86.3,186.2H70.9V79.1h15.4v48.4V186.2z}
		svg{M108.9,79.1h41.6c39.6,0,57,28.3,57,53.6c0,27.5-21.5,53.6-56.8,53.6h-41.8V79.1z M124.3,172.4h24.5c34.9,0,42.9-26.5,42.9-39.7c0-21.5-13.7-39.7-43.7-39.7h-23.7V172.4z}
		svg{M88.7,56.8c0,5.5-4.5,10.1-10.1,10.1c-5.6,0-10.1-4.6-10.1-10.1c0-5.6,4.5-10.1,10.1-10.1C84.2,46.7,88.7,51.3,88.7,56.8z};
	}
}

\newcommand\orcidicon[1]{\href{https://orcid.org/#1}{\mbox{\scalerel*{
				\begin{tikzpicture}[yscale=-1,transform shape]
					\pic{orcidlogo};
				\end{tikzpicture}
			}{|}}}}
			
\newcommand{\Tau}{\mathrm{T}}

\theoremstyle{definition}
\newtheorem{definition}{Definition}
\newtheorem*{remark}{Remark}

\begin{document}
%
%
%
%

\author{David~Vosahlik \orcidicon{0000-0002-5817-8173}\,,
	Tomas~Hanis \orcidicon{0000-0003-1738-2589}
\thanks{D. Vosahlik and T. Hanis are with the Department of Control Engineering, Faculty of Electrical Engineering, Czech Technical University in Prague, Czech Republic. Email: vosahda1@fel.cvut.cz, hanistom@fel.cvut.cz}
\thanks{D. Vosahlik was supported by the Grant Agency of the CTU in Prague, grant No. SGS22/166/OHK3/3T/13.}
}
%
%

\markboth{IEEE TRANSACTIONS ON INTELLIGENT TRANSPORTATION SYSTEMS}%
{Shell \MakeLowercase{\textit{et al.}}: Bare Demo of IEEEtran.cls for IEEE Journals}

\title{Traction Control Allocation Employing Vehicle Motion Feedback Controller for Four-wheel-independent-drive Vehicle}

\onecolumn
\Large{\begin{center}
		\copyright 2023 IEEE.  Personal use of this material is permitted.  Permission from IEEE must be obtained for all other uses, in any current or future media, including reprinting/republishing this material for advertising or promotional purposes, creating new collective works, for resale or redistribution to servers or lists, or reuse of any copyrighted component of this work in other works.
\end{center}}
\normalsize

Accepted to be published in: IEEE Transactions on Itelligent Transportation Systems
\newpage

\newpage
\newpage
\twocolumn

\normalsize

\maketitle

\begin{abstract}
	
  A novel vehicle traction algorithm solving the traction force allocation problem based on vehicle center point motion feedback controller is proposed in this paper. 
  The center point motion feedback control system proposed utilizes individual wheel torque actuation assuming all wheels are individually driven. 
  The approach presented is an alternative to the various direct optimization-based traction force/torque allocation schemes.
  The proposed system has many benefits, such as significant reduction of the algorithm complexity by merging most traction system functionalities into one. 
  Such a system enables significant simplification, unification, and standardization of powertrain control design. 
  Moreover, many signals needed by conventional traction force allocation methods are not required to be measured or estimated with the proposed approach, which are among others vehicle mass, wheel loading (normal force), and vehicle center of gravity location.
  Vehicle center point trajectory setpoints and measurements are transformed to each wheel, where the tracking is ensured using the wheel torque actuation. The proposed control architecture performance and analysis are shown using the nonlinear twin-track vehicle model implemented in Matlab $\&$ Simulink environment. The performance is then validated using high fidelity FEE CTU in Prague EFORCE formula model implemented in IPG CarMaker environment with selected test scenarios. Finally, the results of the proposed control allocation are compared to the state-of-the-art approach.

\end{abstract}


\IEEEpeerreviewmaketitle

\section{Introduction}
 \label{sec:intro}
 
 The rising demand for higher efficiency, safety, and advanced functionality of the modern vehicle traction system gives rise to often overly complicated solutions, both from a mechatronics and software engineering point of view. The latter becomes of even greater importance these days. 

 The task of vehicle traction control is difficult due to unavailable directly measured or detected wheel traction force and traction force allocation task connected to the over-actuated system (up to four wheels are driven independently, and usually all are individually braked). 
 The control allocation problem is commonly solved via a direct optimal distribution of traction force, torque, or slip ratio over the wheels, as shown in the traction control survey paper \cite{Ivanov2015}. 
 The control allocation solved via traction force constrained minimization is proposed in \cite{Guo2018}. Similarly, in \cite{Maeda2012a}, the authors propose the solution via a minimum least-squares formulation minimizing the slip ratio for each wheel. Analogous optimal traction force allocation with different optimization criteria is shown in \cite{DeCastro2012}. 
 In \cite{Park2015}, the authors are augmenting the same approach as in \cite{Maeda2012a} with lateral dynamics and lateral force distribution. The combination of longitudinal and lateral force is minimized. Another approach was shown in \cite{Tian2019}, where the individual wheel traction force reference is computed based on the wheel and vehicle parameters, commanded vehicle yaw rate, and longitudinal acceleration. 
 Another optimization-based approach to control allocation was shown in \cite{Li2020}, where the authors propose a neural network and sliding mode controller to solve the control allocation.
 In \cite{Hu2019}, the authors are even neglecting the tire model while assuming the wheel torque will be losslessly transformed into the traction force. Such an approximation is acceptable for low slip ratios but is not valid for dynamic maneuvers where the slip ratios are higher, and the tire model has a significant impact. A similar approach is presented in \cite{Ge2011} and \cite{Amato2019}, where the authors assume that the ideal force can be computed on the level of vehicle motion. 
  
 The traction force of a particular wheel can be considered as a monotonic function of the wheel slip ratio for the stable operating range of the slip curve considering its typical shape (see \ref{fig:unstablePacejka}). Therefore, the traction force and slip ratio allocation are, in principle, following the same idea of the traction force distribution. 
 The environment parameters like wheel normal force/load, tire-to-road interface friction coefficient, and reduction of longitudinal wheel capacity due to wheel lateral force can be expressed as slip curve scaling and its parameters variation (see \cite{Pacejka2002}).

 \subsubsection*{Novel vehicle motion feedback allocation method}
 A novel alternative to the control allocation methods is proposed in this paper, which is the main contribution. The driver-to-vehicle interface is based on a common assumption -- a self-driving planning algorithm is commanding a vehicle speed and yaw rate or a driver commanding vehicle acceleration and yaw rate at the vehicle's Center Point (CP). The CP is where the vehicle motion references (like velocity, acceleration, yaw rate, etc.) are to be tracked and where the respective signals are measured. The CP is located by a designer arbitrarily depending on the particular vehicle. 
 The general motion of a rigid body object in a plane is fully parameterized by its longitudinal, lateral, and angular speed/acceleration of the rigid body's arbitrary point. Therefore, the vehicle CP variables would be used for a driver-to-vehicle interface without loss of generality. Furthermore, the CP position could be selected arbitrarily. The vehicle measurement points could be distributed all over the vehicle. However, it is assumed that all vehicle states are acquired (measured or estimated) at CP.
 The vehicle velocity and acceleration state variables, references, and measurements can be directly, uniquely, and unambiguously transformed from vehicle CP to wheel pivot points (for details, see eq.~\ref{eq:transform}). 
 The wheel-level control system could be designed in a centralized or distributed fashion.
 The main focus of this paper is the traction control allocation. 
 However, a similar control system proposed here can also be drawn for lateral dynamics.
 The paper's main contribution is a novel approach to control allocation. The control structure presented has many benefits over the traction systems based on a direct force/slip ratio allocation. The benefits are presented and discussed in the \ref{sec:benefits}.
 
 \subsubsection*{Assumptions}
 All the necessary signals for the proposed system are assumed to be available (measured or estimated), generally at any arbitrary point, however, let us select following:
 \begin{itemize}
     \item At the vehicle center: vehicle longitudinal velocity \& acceleration and vehicle angular rates (like yaw rate, etc.).
     \item At each wheel: wheel RPM, wheel traction torque, and wheel steering angle.
 \end{itemize}
 The paper structure is as follows. First, the transparent, easy to analyze, open, easy to modify, and still high fidelity nonlinear model implemented in Simulink is presented in \ref{sec:model}. The model is very beneficial for showing and verifying particular proposed controller properties and functionalities. Afterward, the proposed control system with its analysis and simulations showing the performance using the  Matlab \& Simulink non-linear model is presented. Finally, the IPG CarMaker high fidelity EFORCE formula validation model is introduced to validate the controller and compare it to the state-of-the-art approach presented in \cite{Maeda2012a}. The CarMaker model is parameterized using real CTU student EFORCE formula measurements.

\section{High fidelity mathematical model}\label{sec:model}
The nonlinear twin-track model adopted from \cite[Chapter~11]{Schramm2014} and \cite[Chapter~2]{vita_diplomka} is considered for particular controller functionality validation $\&$ verification purposes. Implementation of the model is available in git \cite{TwinTrack}. The model has multiple coordinate systems (CS). Superscripts are used for the expression of the CS. The CS used in the model (see \ref{fig:CtrlCS}) are presented below:
\begin{itemize}
    \item \textbf{vehicle body-fixed} has its origin at vehicle CP. All vehicle measurements and driver commands are assumed to be located in the point without loss of generality. For the system, the superscript $v$ is used.
    \item \textbf{wheel pivot-fixed} has its origin in the wheel pivot point oriented the same as vehicle body. The superscript $b_i$ is used for the system, where $i$ stands for $i$-th wheel. Orientation of the coordinate system is the same as in the vehicle body-fixed CS. The wheel pivot-fixed CS is a translation of the vehicle body-fixed CS along the vector $\vect{r_i}$.
    \item \textbf{wheel-fixed} has its origin in the center of the wheel. It is bound to the wheel (including the orientation). The superscript $w_i$ is used for this system, where $i$ stands for $i$-th wheel. The wheel-fixed CS is rotated wheel pivot-fixed CS by the wheel steering angle $\delta_i$.
\end{itemize}
The mathematical model consists of four fundamental parts. 
\begin{itemize}
    \item Vehicle nonlinear rigid body, see \ref{sec:Single_trac_model}
    \item Suspension, to model load transfer
    \item Tire-to-road interface, see \ref{sec:tireInterface}
    \item Drivetrain model including wheel, see \ref{sec:wheel_model}
\end{itemize}

Vehicle suspension is implemented in the model but is not presented here as it is not a key part of the model. Interested reader is directed to \cite{Schramm2014,TwinTrack} for further details.
\subsection{Vehicle nonlinear dynamics model}\label{sec:Single_trac_model}
 While describing the model, the CP is set to be exactly in the vehicle Center of Gravity (CG) without loss of generality to simplify the model equations  derivation. Nevertheless, the CG is located rearwards from the geometric center of the vehicle.

 \begin{figure}[h] 
    \centering
    \includegraphics[width=1\linewidth]{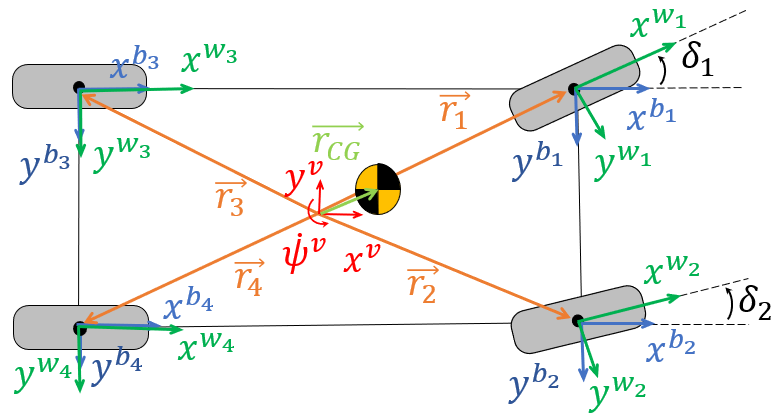}
    \caption{Coordinate systems considered. $xy$ plane view.}
    \label{fig:CtrlCS}
\end{figure}

The twin-track model state variables are $\vect{x} = [\vect{v}^{v}, \vect{\Omega}^v, \omega_1, \dotsi, \omega_N]$, where $\vect{v}^{v}$ is vehicle velocity vector at CG (vehicle body-fixed CS), $\vect{\Omega}^v$ is vehicle body angular rates vector at CG (vehicle body-fixed CS), $\omega_{i}$ where $i \in \{1,2,3,4\}$ represents $i$-th wheel are wheel angular velocities (wheel-fixed CS). 
The twin-track model inputs are $\vect{u}=[\tau_{\text{ref}\,1}, \dotsi,\tau_{\text{ref}\,N}, \delta_1, \delta_2]$, where $\tau_{\text{ref}\,i}$ is commanded torque for $i$-th wheel axis ($i = \{1,2,3,4\}$ stands for each wheel number) and $\delta_1, \delta_2$ are steering angles of the front wheels. 
The twin-track model coordinate system is presented in \ref{fig:CtrlCS}. Vehicle yaw rate $\Omega_z^v$ is also noted as $\dot{\psi}^{v}$.

The steering angles $\delta_i$ are constrained to 

    \begin{equation}\label{eq:deltaConst}
    \delta_1 = \delta_2 \overset{!}{=} \delta,
    \end{equation}
where a driver commands $\delta$. The Ackerman steering is neglected for the sake of the mathematical nonlinear model calrity. However, it is implemented in the high fidelity CarMaker model for final validation of the proposed control system.

The dynamics of the twin-track rigid body is based on Newton-Euler equations as

\begin{subequations} \label{eq:nonLinSystem}
    \begin{equation}\label{eq:NewtonEq}
        m_v ( 
        \dot{\vect{v}^v} + \vect{\Omega}^v \times \vect{v}^v)
        =  \vect{F}^v,
    \end{equation}
    \begin{equation}\label{eq:EulerEq}
        \mathbf{\Theta}_v \cdot \dot{\vect{\Omega}}^v + \vect{\Omega}^v \times (\mathbf{\Theta}_v \cdot\vect{\Omega}^v) 
        = \vect{\Tau}^v,
    \end{equation}
\end{subequations}
where $\vect{F}^v$ is the resulting force acting on the vehicle at CG, $\vect{\Tau}^v$ is the resulting torque acting on the vehicle at CG, $m_v$ is vehicle mass, and $\mathbf{\Theta}_v$ is the tensor of vehicle moment of inertia at CG. Then,

\begin{align}
    \vect{F}^v &= \sum_{i=1}^{4} T_v^{b_i}\cdot \vect{F}^{b_i} - \vect{F}_{res} + \vect{F}_g^v,\\
    \vect{\Tau}^v &= \sum_{i=1}^{4} \vect{r}_i^v \times \vect{F}^{b_i} + \vect{r}_{res}^v \times \vect{F}_{res},\\
    \vect{F}_{res} &= \frac{1}{2}c_{res} \rho A \sqrt{\left(v^{v}_x\right)^2 + \left(v^{v}_y\right)^2}
    \begin{bmatrix}
    v_x^{v}\\v_y^{v}\\0
    \end{bmatrix},
\end{align}

where  $\vect{F}^{b_i}$ is vector of forces generated by $i$-th wheel (wheel pivot-fixed CS) and transformned to vehicle body-fixed frame via trivial identity transformation $T_v^{b_i}$, $\vect{F}_g^v$ is gravitational force (vehicle body-fixed CS), $\vect{F}_{res}$ is combination of all resistant and aerodynamic forces acting on the vehicle at center of pressure (CPr), $\vect{r}_{res}^v$ is position vector of the CPr (vehicle body-fixed CS),   $\vect{r}_i^v$ is position of $i$-th wheel, $c_{res}$ is aerodynamic drag constant, $\rho$ is air density and $A$ is the equivalent frontal area.

 \subsection{Drivetrain and wheel model}
\label{sec:wheel_model}
The drivetrain and wheel level model is introduced in this section. The system dynamics is characterized as
\begin{equation}
J_{i}\cdot\dot{\omega}_{i} = \tau_i + F_{x}^{w_i}\cdot r_{w_i} + \tau_{res,i}(\omega_i),
\label{eq:wheel_dynamics}
\end{equation}
where $J_{i}$ is wheel moment of inertia along the wheel shaft ($y$) axis (including also whole drivetrain rotational inertia), $\omega_{i}$ is wheel angular speed along the wheel shaft axis, $\tau_i$ is wheel drive/brake torque, $F_{x}^{w_i}$ is longitudinal traction force generated by particular wheel and represents vehicle dynamics, $\tau_{res,i}(\omega_i)$ are all-wheel and e-motor losses combined as function of wheel speed and $r_{w_i}$ is $i$-th wheel effective radius. The road condition, wheel load/normal force, and the effect of wheel lateral force are modeled by the reduction of wheel longitudinal traction force $F_{x}^{w_i}$. The reduction is modeled by the wheel longitudinal slip curve (\ref{fig:slip_curve}) and traction ellipse shape (\ref{fig:tractionellipse}). The dependency is sketched as

\begin{equation}
F_{x}^{w_i}=f(\lambda_i, F_{z}^{w_i}, \mu_i, F_{y}^{w_i}).
\label{eq:Fx_function}
\end{equation}

The actuator model is introduced to capture the powertrain system delays, e.g., CAN bus communication delay, and neglected dynamics (e.g., inverter dynamics). The first-order actuator model is introduced into the vehicle model (see eq.~\ref{eq:torque_dynamics}) in order to shape the actual e-motor torque $\tau_i$ that is used in the wheel dynamics (see eq.~\ref{eq:wheel_dynamics}). The model takes the commanded e-motor torque $\tau_{\text{ref},i}$ and shapes the actual torque $\tau_i$ according to 

\begin{equation}
\dot{\tau_i} = \dfrac{1}{T}(\tau_{\text{ref},i} - \tau_i).
\label{eq:torque_dynamics}
\end{equation}

Usually, the system delays reach up to ten milliseconds. The worst-case scenario time constant, namely, $T = 10ms$, is assumed in this paper and implemented in the model. The traction and brake torque control and blending may be implemented as proposed in \cite{Pugi2020}.

\subsection{Tire-to-road interface}
\label{sec:tireInterface}

The wheel forces generated at tire-to-road interface are commonly described using slip variables (longitudinal slip ratio $\lambda$ and slip angle $\alpha$, see eq.~\ref{eq:slip-ratio} and \ref{eq:slip_angle}). The Pacejka magic formula for longitudinal and lateral direction is used for tire slip curve model (see \cite{ Schramm2014}). Finally, the traction ellipse is employed to bind the wheel's longitudinal and lateral traction properties (see \cite{Pacejka2002,FricEllipse} for further details.)

The wheel longitudinal traction force is governed by the wheel slip ratio $\lambda_i$ defined by eq.~\ref{eq:slip-ratio}. 

\begin{equation}
\lambda_i = \frac{\omega_{i} \cdot r_{w_i} - v_{x}^{w_i}}{max(|\omega_{i}|\cdot r_{w_i},|v_{x}^{w_i}|)},
\label{eq:slip-ratio}
\end{equation}
where $v_{x}^{w_i}$ is $i$-th wheel hub longitudinal speed (wheel-fixed CS), $\omega_{i}$ is $i$-th wheel angular speed and $r_{w_i}$ is $i$-th wheel effective radius.
The dependency of the longitudinal force generated by a particular wheel $F^{w_i}_{x}$ on the wheel slip ratio and tire-interface variable is represented by the slip curve and simplified Pacejka magic formula (see \cite{Pacejka2002}) defined as
\normalsize

\begin{figure} 
    \centering
    \includegraphics[width=0.5\linewidth]{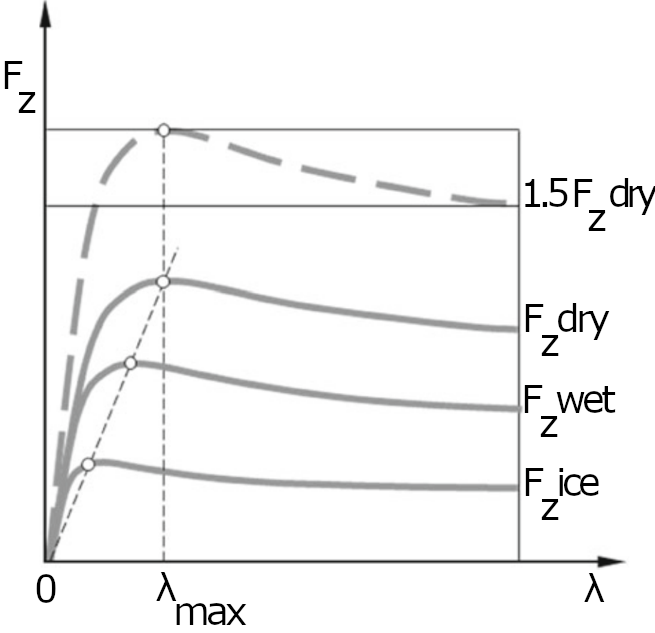}
    \caption{Longitudinal slip curve. Adopted from \cite{Schramm2014}. The same shape is observable for negative slip ratios generating negative/braking traction force.}
    \label{fig:slip_curve}
\end{figure}

\small
\begin{equation}\label{eq:pacejka_long}     
            \overline{F^{w_i}_x} = \mu_i F_z^{w_i}D\sin(C  \arctan (B \lambda - E (B \lambda - \arctan(B \lambda)))),
\end{equation}
\normalsize
where $\mu_i$ is road friction coefficient, $F_{z}^{w_i}$ is normal force acting on the wheel, $B$, $C$, $D$, $E$ are Pacejka's shaping coefficients and $\overline{F^{w_i}_x}$ is force generated by the wheel in the $x$ direction, the traction force (see \ref{fig:slip_curve}). 

\begin{figure} 
    \centering
    \begin{subfigure}[b]{0.47\columnwidth}
        \includegraphics[width=\linewidth]{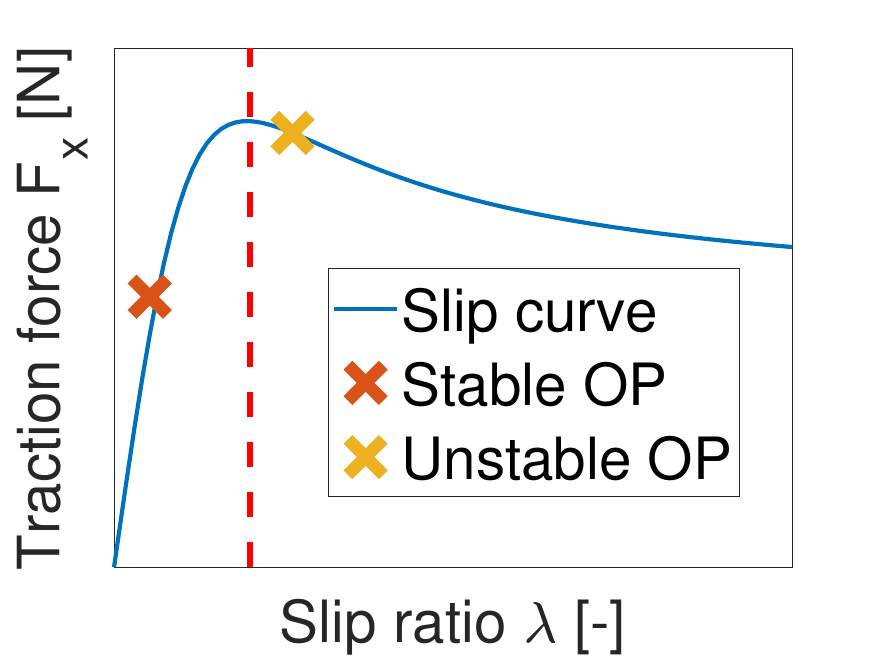}
        \vskip7mm
        \caption{Stable and unstable Operating Point (OP).}
        \label{fig:unstablePacejka}
    \end{subfigure}
    \centering
    \begin{subfigure}[b]{0.5\columnwidth}
        \includegraphics[width=\linewidth]{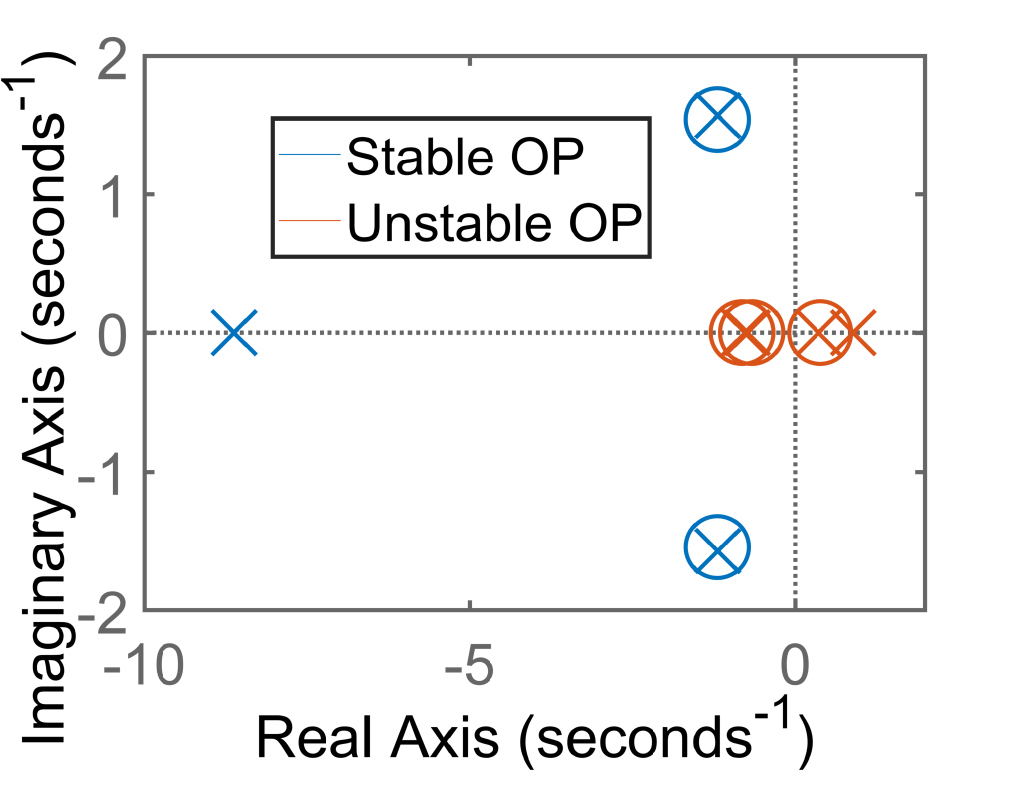}
        \caption{Root locus of minimum realization stable and unstable OP linearized vehicle model.}
        \label{fig:rlNoCtrl}
    \end{subfigure}
    \caption{The stable and unstable slip curve OPs.}
\end{figure}

The slip curve can be divided into two regions depending on the slip curve slope -- stable with a positive slip curve slope and unstable, where the slip curve slope is negative.
The stable region Operating Point (OP) linearization of the wheel dynamics \ref{eq:wheel_dynamics} yields a stable system and vice versa.
An example of various OPs linearization is shown in the \ref{fig:unstablePacejka}. The minimum realization root locus of the whole vehicle model linearization in those points is shown in \ref{fig:rlNoCtrl}.

It is essential to notice the dependency of the resulting traction force on wheel normal loading $F_{z}^{w_i}$ and road friction coefficient $\mu_i$. A new artificial tire interface variable is introduced to bind all these dependencies together and have one convenient-to-use variable. Such simplification is neglecting the slip curve maximum shift due to tire-to-road interface change, combined loading changes or lateral slip effects (see \ref{fig:slip_curve}).

\begin{definition}{~}\label{def:eps}
    Tire-interface variable $\epsilon_i$ is defined for an $i$-th wheel as
    
        \begin{equation}  \label{eq:eps}      
            F^{w_i}_x = \epsilon_i \cdot \overline{F^{w_i}_x},~~~\epsilon_i \in {\rm I\!R}^+.
        \end{equation}
         
    
\end{definition}

\begin{remark}
    \textit{If $\epsilon_i$ = 1 then the equation~\ref{eq:pacejka_long} becomes standard Pacejka magic formula \cite{Pacejka2002}.}
\end{remark}

The wheel lateral traction force is governed by the wheel slip angle $\alpha_i$, defined as
\begin{equation}
\alpha_{i} = -\arctan(\frac{v^{w_i}_{y}}{|v^{w_i}_{x}|}),
\label{eq:slip_angle}
\end{equation}
where $v^{w_i}_{x}$ is $i$-th wheel hub longitudinal speed and $v^{w_i}_{y}$ is $i$-th wheel hub lateral speed (wheel-fixed CS).

The dependency of the lateral force generated by a particular wheel $F^{w_i}_{y}$ on such wheel slip angle is represented by the slip curve and Pacejka magic formula with the same formulation as presented in eq.~\ref{eq:pacejka_long}, however with slip variable $\alpha_i$ instead of $\lambda_i$ and with different set of parameters representing lateral tire-to-road interface properties \cite{Pacejka2012}.

Finally, the traction ellipse is introduced to bind the longitudinal and lateral wheel traction capacity. The traction ellipse represents the bounded friction force generated by the tire-to-road contact patch in combined longitudinal and lateral motion. A combined slip occurs when the vehicle accelerates or brakes in a cornering maneuver. A tire cannot generate a combined traction force (comprising of lateral and longitudinal components) greater than the $F_{z}^{w_i}\cdot\mu_i$, where the $F_{z}^{w_i}$ is the normal force. That restriction is expressed by the friction ellipse (also called Kamm's circle):
\begin{equation}
F_{combined}^{w_i} = \sqrt{\frac{\left(F_x^{w_i}\right)^2}{(\mu_i D_x)^2}+\frac{\left(F_y^{w_i}\right)^2}{(\mu_i D_y)^2}} \leq \mu_i F_z^{w_i},
\end{equation} 
where $D_x$ and $D_y$ are longitudinal and lateral Pacejka magic formula $D$ parameters.

\begin{figure}[h] 
    \centering
    \begin{subfigure}[b]{0.45\columnwidth}
        \includegraphics[width=\linewidth]{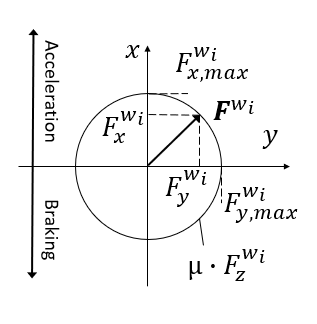}
    \caption{Wheel traction ellipse.}
    \label{fig:tractionellipse}
    \end{subfigure}
    \begin{subfigure}[b]{0.48\columnwidth}
        \includegraphics[width=\linewidth]{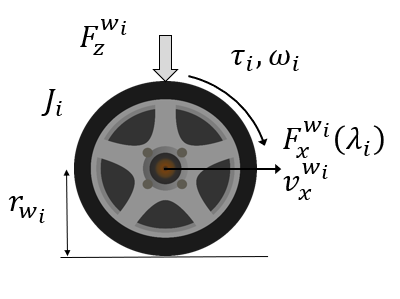}
    \caption{Wheel model}
    \label{fig:wheelmodelsideview}
    \end{subfigure}
    \caption{Wheel model.}
\end{figure}

The friction ellipse is shown in the \ref{fig:tractionellipse}. Its implementation is adopted from \cite{Efremov2019} which refers to \cite{FricEllipse, Pacejka2002}. The equations are not repeated here again for the sake of brevity but can be found in already mentioned \cite{FricEllipse,Efremov2019} or in the acopmanying technical report \cite{FormulaGit}.

\section{Vehicle motion feedback control allocation architecture}
\label{sec:controller}

\begin{figure*}[htb]
    \centering
    \includegraphics[width=1\linewidth]{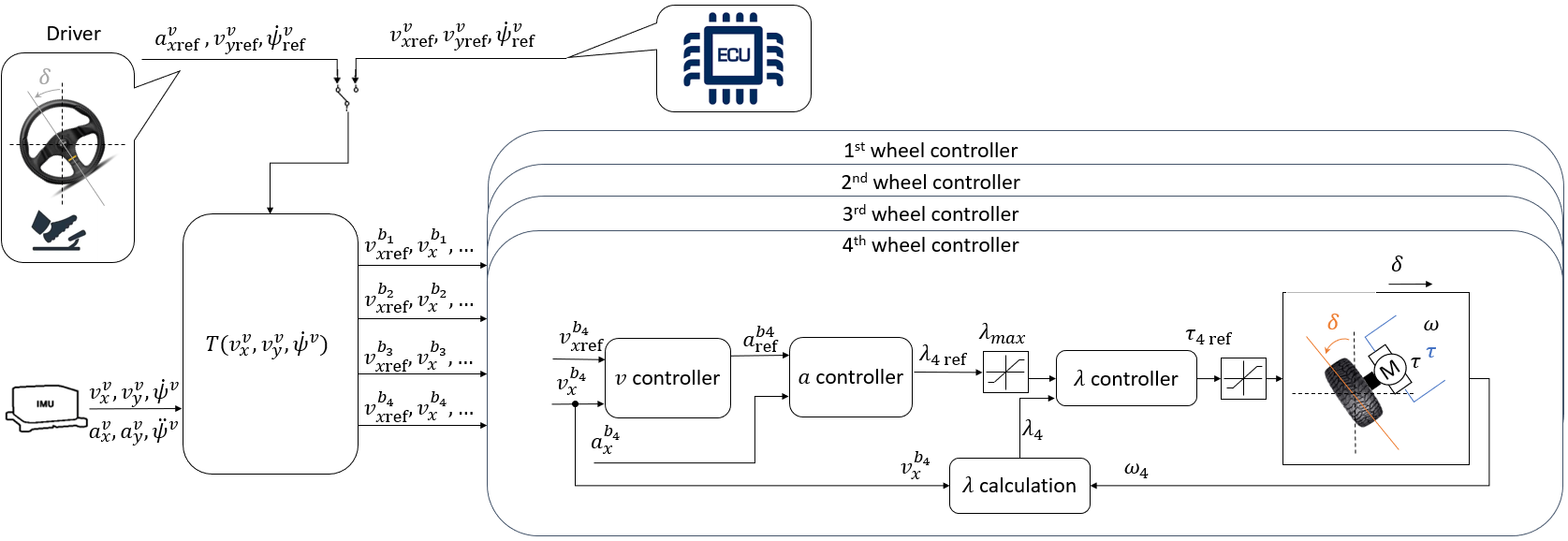}
    \caption{Proposed traction control system architecture.}
    \label{fig:FAFT_arch}
\end{figure*}

The main contribution of this paper is the proposed control architecture providing a system-level design approach to wheel level longitudinal dynamics control. 
The main idea is to use transformation \ref{eq:transform} which transforms vehicle states (speed or acceleration), both references and measurements, onto individual wheels. Whereas the direct optimization-based methods used in \cite{Guo2018, Maeda2012a, DeCastro2012} determine the individual wheel slip ratio or traction force utilizing optimization to solve the ambiguity of control allocation task. 
Then, the vehicle level variables are controlled at the wheel level using any desired control strategy.

The transformation of vehicle level signals onto the wheel level is defined as

\begin{equation}\label{eq:transform}
    \vect{v}^{b_i}_{~\text{ref}} = T(\vect{v}^{v}_{~\text{ref}}, \vect{\Omega}_{~\text{ref}}^{v}) = \vect{v}^{v}_{~\text{ref}} + \vect{\Omega}_{~\text{ref}}^{v} \times \vect{r}^{v}_{~i} + \vect{\gamma}(\vect{\Omega}_{~\text{ref}}^{v}),
\end{equation} 
where $\vect{v}^{v}_{~\text{ref}}$ is commanded velocity vector at vehicle CP,  $\vect{\Omega}_{\text{ref}}^{v}$ are the desired roll, pitch, and yaw rates at CP, $\vect{r}^{v}_{~i}$ is position of $i$-th wheel and $\vect{v}^{b_i}_{~\text{ref}}$ is velocity reference signal for $i$-th wheel. As the scope of this work is mainly wheel longitudinal dynamics, only the $v_{x\,\text{ref}}^{b_i}(v_x^v, \dot{\psi}^v)$ as function of vehicle velocity $v_x^v$ and yaw rate $\dot{\psi}^v$ is extracted from the results of the transformation and further controlled. However, an extension to the lateral speed $v_{y\,\text{ref}}^{b_i}$ tracking is straightforward.

The $\vect{\gamma}$ function is augmenting the pure physical transformation \ref{eq:transform} introducing lateral vs longitudinal dynamics preference feature. The $\vect{\gamma}$ function is derived out of the vector cross-product definition and is defined as 
\begin{equation}\label{eq:transformAug}
    \vect{\gamma}(\vect{\Omega}^{v}) = (\Gamma \cdot (-\dot{\psi}^v\cdot r^{v}_{y~i}),~0,~0),
\end{equation}
where $\Gamma \in {\rm I\!R}^+_0$ is the tuning parameter. If $\Gamma = 0$ the transformation \ref{eq:transform} preserves the original physics-based meaning.
$\Gamma$ value amplifies the effect of yaw rate on the resulting $v_x^{b_i}$ in \ref{eq:transform}. Yaw rate $\dot{\psi}^v$ is this way preferred over the longitudinal velocity $v_x^{v}$ tracking. The designer can appropriately tune $\Gamma$ particular value to the specific vehicle lateral to longitudinal dynamics control preference.
Physically, $\Gamma$ could be understood as a parameter artificially inflating the vehicle width.

The wheel level layer control laws -- in this paper's particular case, velocity, acceleration, and slip ratio tracking -- are presented here for the sake of completeness. However, it is essential to say it could be replaced by any other suitable control mechanism providing the functionality.

\subsection{Drive-by-human and self-driven controller hierarchy} \label{sec:drive-by-human}

The hierarchical architecture of the wheel level control layer is presented in the \ref{fig:FAFT_arch}. The transformation \ref{eq:transform} is general and can be employed in the case of self-driving (automated) or human-driven vehicles. The trajectory planning algorithms commonly generate vehicle velocity and yaw rate profiles to follow, which perfectly fits the presented transformation \ref{eq:transform}. In that case, the transformation \ref{eq:transform} is directly used. 

However, the control system proposed can also be employed in the human driver-driven vehicle, where vehicle acceleration and yaw rate seem to be more suitable variables to control. In this case the transformation \ref{eq:transform} has to be modified as
\begin{equation}\label{eq:transformDriver}
    \vect{v}^{b_i}_{~\text{ref}} = T(0, \vect{\Omega}_{~\text{ref}}^{v}) = 0 + \vect{\Omega}_{~\text{ref}}^{v} \times \vect{r}^{v}_{~i} + \vect{\gamma}(\vect{\Omega}_{~\text{ref}}^{v}).
\end{equation} 
Then, the reference for the wheel acceleration controller $a^{b_i}_{x~\text{ref}}$ is determined as
\begin{equation}\label{eq:accelerationRef}
    a^{b_i}_{x~\text{ref}} = a^{v}_{x~\text{ref}} + a^{b_i}_{\text{lat}},
\end{equation}
where $a^{b_i}_{\text{lat}}$ is the acceleration commanded by the wheel pivot point velocity controller representing the corrective action for $\vect{\Omega}_{~\text{ref}}^{v}$ tracking and $a^{v}_{x~\text{ref}}$ is the driver generated CP acceleration command.

\subsection{Wheel controller hierarchy} \label{sec:longVelo}
The wheel level control layer can be replaced by any suitable control system and is here presented for controller completeness. The controller can be designed either in a centralized or distributed manner.

The wheel level part of the control system is composed of three hierarchically connected controllers. Namely, the wheel longitudinal slip ratio $\lambda_i$, wheel pivot point acceleration $a_{x}^{b_i}$ and wheel pivot point velocity $v_{x}^{b_i}$ controllers. The wheel pivot point velocity $v_{x}^{b_i}$ is commanded from the transformation \ref{eq:transform} and employs wheel pivot point acceleration $a_{x~\text{ref}}^{b_i}$ command as manipulated variable. Then, the wheel pivot point acceleration $a_{x}^{b_i}$ is controlled via the  $\lambda_{i~\text{ref}}$ command. Last, the wheel slip ratio $\lambda_i$ is controlled manipulating the wheel torque $\tau_{i~\text{ref}}$. All the controllers were designed using continuous time design techniques and then discretized with sampling time $T_s = 0.01s$. The sampling time was chosen so that the control system proposed is easily deployable on the embedded HW used in the vehicles.

\begin{figure*}
    \centering
    \includegraphics[width=0.85\linewidth]{"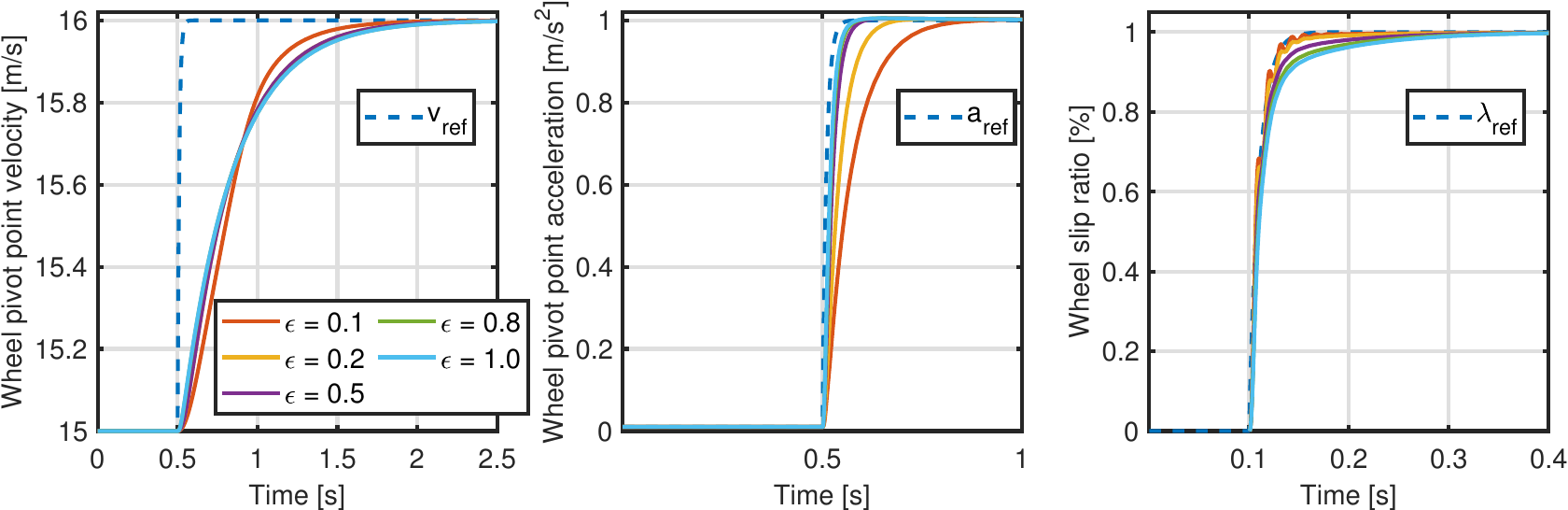"}
    \caption{Wheel level individual control loops step responses. The step responses are from left to right: Wheel pivot point velocity tracking, wheel pivot point acceleration tracking, and slip ratio tracking.}
    \label{fig:accel_step}
\end{figure*}

The wheel level acceleration and velocities are hard and costly to measure. Therefore, only the CP measurements (vehicle body states measurements) are used. The wheel level velocities' and accelerations' measurements are computed via the transformation \ref{eq:transform} (as sketched in the \ref{fig:FAFT_arch}). To sum up, the building blocks, from the inner to the outer loop, are:

\begin{itemize}
    \item \textbf{$\lambda_i$ tracking} - inner/core layer providing wheel slip ratio tracking functionality via torque $\tau_{i~\text{ref}}$ manipulation.
    \item \textbf{$a_{x}^{b_i}$ tracking} - middle layer provides wheel pivot point acceleration $a_{x}^{b_i}$ tracking functionality via $\lambda_{i~\text{ref}}$ manipulation.
    \item \textbf{$v_{x}^{b_i}$ tracking} - most outer layer controlling wheel pivot velocity via $a_{x~\text{ref}}^{b_i}$ manipulation.
\end{itemize} 

Step responses of the individual loops are presented in \ref{fig:accel_step}.

\subsubsection{$\lambda$ tracking}
\label{sec:lambda_tracking}
 Wheel slip ratio (assuming stable OP) can be physically interpreted as a proportional part of the available traction force. It is even more apparent once the bilinear slip curve approximation and tire stiffness $c_\lambda$ is used as 
\begin{equation}\label{eq:linearApprox}
    F_x = F_zc_\lambda \lambda.
\end{equation}
The slip ratio and wheel angular speed control problems are already solved in many works, e.g., in \cite{Jia2018}, the authors are proposing a piece-wise linear feedback controller that controls the slip ratio $\lambda$ and traction force for each wheel, respectively. A similar problem is solved in \cite{Ringdorfer2011}. The slip ratio seems to be the best candidate for the controlled variable of any traction system. The idea of $\lambda$-control is not new and such control strategies are common in railroad vehicles (see \cite{Lehtla2008}), but could also be found in current industrial R\&D automotive projects such as \cite{ECV1000}. All the necessary safety/economy functionality of the wheel-level layer controllers is preserved for the proposed controller. 

The $\tau_{\text{ref},i}$ value is generated based on demanded $\lambda_{i~\text{ref}}$ value, wheel hub longitudinal speed $v_{x}^{w_i}$, and wheel speed $\omega_i$. 
At first, the measured values at vehicle CP are transformed using eq.~\ref{eq:transform} to the wheel pivot point. 
Then, the measured slip ratio is computed using the eq.~\ref{eq:slip-ratio}.
The controller of the slip ratio was designed with linear techniques, namely root locus, and then fine-tuned on the nonlinear model presented in \ref{sec:model}. A full PID controller was designed on linearized vehicle model having OP with nonzero slip ratio ($\lambda_{OP} = 0.05$). 

\begin{figure}[h] 
    \centering
        \includegraphics[width=0.9\linewidth]{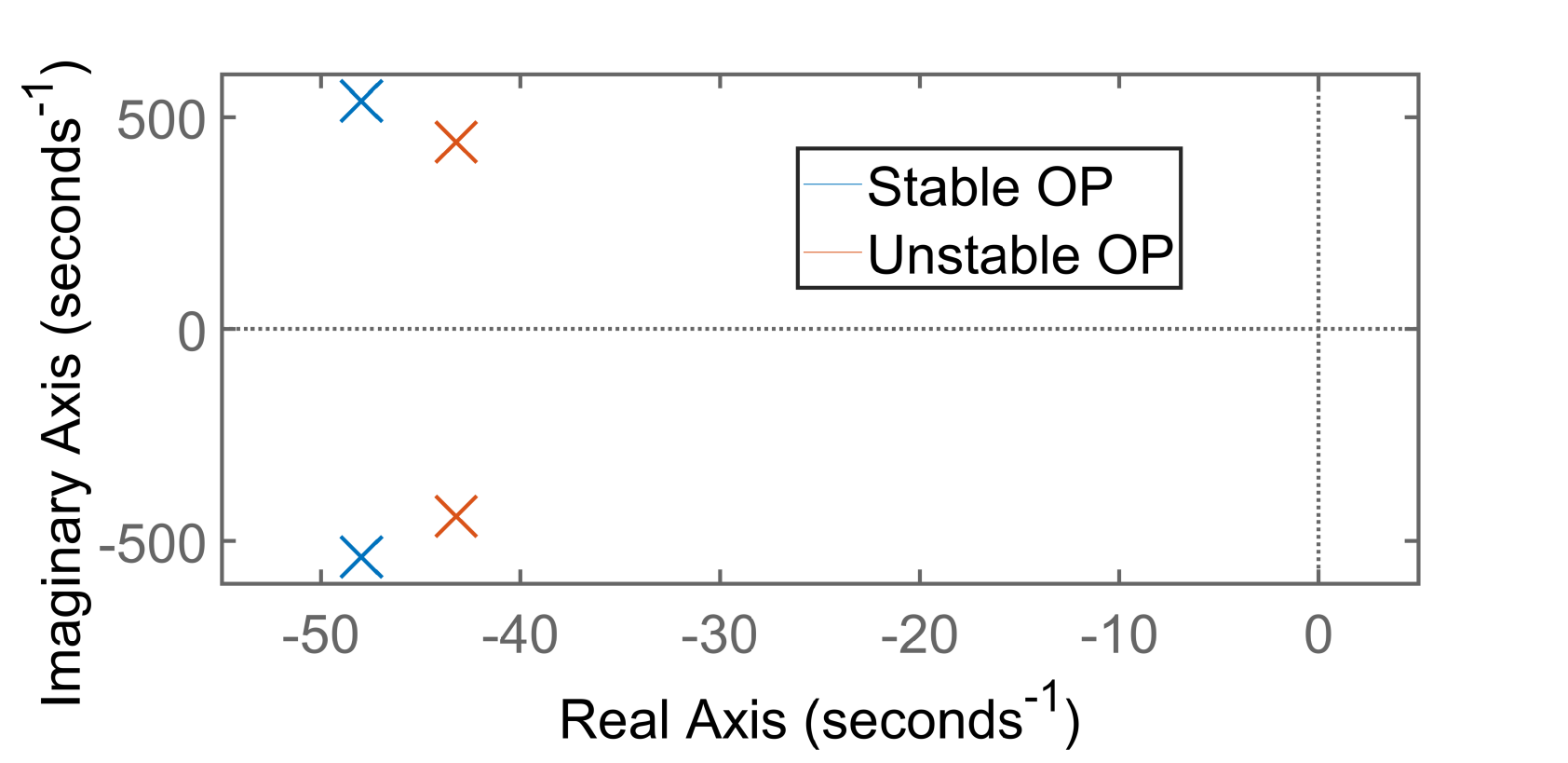}
    \caption{Compensated stable and unstable OP linearized system root locus.}
    \label{fig:rlCtrl}
\end{figure}

The stable and unstable slip curve OPs (see \ref{fig:unstablePacejka}) linear stability analysis is shown in the \ref{fig:rlCtrl}. Two OPs minimum realizations of controlled vehicle dynamics linearization are shown there. The controller was designed for the stable OP (positive slip curve slope). Unstable OP (negative slip curve slope) stabilized with the same discrete time ($T_s = 0.01s$) PID controller designed for the stable OP is also shown in \ref{fig:rlCtrl}.

It is necessary to mention the control of slip ratio $\lambda$ is possible only when the wheel pivot point has nonzero velocity $v_{x}^{b_i} > \zeta; \zeta > 0$ to enable the slip ratio calculation (see \ref{eq:slip-ratio}). Therefore, the control policy presented, is applicable only for vehicle velocities higher than some threshold, e.g., the vehicle velocity $|\vect{v^v}| > 2 m/s$.

\subsubsection{$a_{x}^{b_i}$ tracking - "$\lambda$-reference generation"}
\label{sec:accel_tracking}
This layer is responsible for $a_{x,\text{ref}}^{b_i}$ (acceleration at $i$-th wheel pivot point) setpoint tracking. The input-output linearization of the nonlinear model, including the $\lambda$ controller, was used in the continuous time PI controller design.
The PI controller was discretized with $T_s = 0.01s$.
The $\lambda_{max}$ limit is the algorithm parameter and might be selected based on slip curve estimation as the value of slip ratio for maximal wheel traction (see \cite{Seyedtabaii2019}). However, in this paper it is implemented constant.

\subsubsection{Traction limits}\label{sec:tracLim}
Vehicle yaw rate tracking shall be preferred over acceleration/deceleration tracking when the vehicle is almost at its traction limits. When any of the commanded slip ratio $\lambda_{i,~\text{ref}}$ is close to saturation $\lambda_{\text{max}}$, the $a^{b_i}_{x~\text{ref}}$ reference is, as opposed to the normal operation (see \ref{eq:accelerationRef}), computed as
\begin{align}
    a^{b_i}_{x~\text{ref}} &= (a^{v}_{x~\text{ref}} - a_{\text{lat}, \text{max}}) + a^{b_i}_{\text{lat}}, \\
    a_{\text{lat}, \text{max}} &= \text{max}(a^{b_i}_{\text{lat}}),~ i \in (1,2,3,4).
\end{align}
This way, the acceleration ensuring the vehicle yaw rate tracking will always be preferred over the longitudinal acceleration in limiting cases.
It is done in the same manner as the self-driven vehicle controller case.

\subsubsection{$v_{x}^{b_i}$ tracking - "$a_{x}^{b_i}$-reference generation"}
\label{sec:vel_tracking}
The wheel pivot point velocity $v_{x}^{b_i}$, representing vehicle states, is transformation of vehicle longitudinal speed $v_{x}^{v}$ , lateral speed $v_{y}^{v}$ and cornering angular speed $\dot\psi^{v}$ measurements at vehicle CP into the wheel pivot point CS (see eq.~\ref{eq:transform}).
Reference speed for each wheel is also derived from this transformation. This is then tracked by P controller generating $a_{x,\text{ref}}^{b_i}$ demand. The nonlinear model's input-output linearization, including the acceleration controller, was used for the P controller design.


\section{Simulations and results}\label{sec:simulation}
The simulation-based experiments are executed to show the system's performance. First, the Matlab \& Simulink simulations using the nonlinear model from \ref{sec:model} are presented to provide better insight into the controller features. Next, a student formula IPG CarMaker high fidelity validation model simulation is presented. 

\begin{figure*}
    \centering
    \includegraphics[width=\linewidth]{"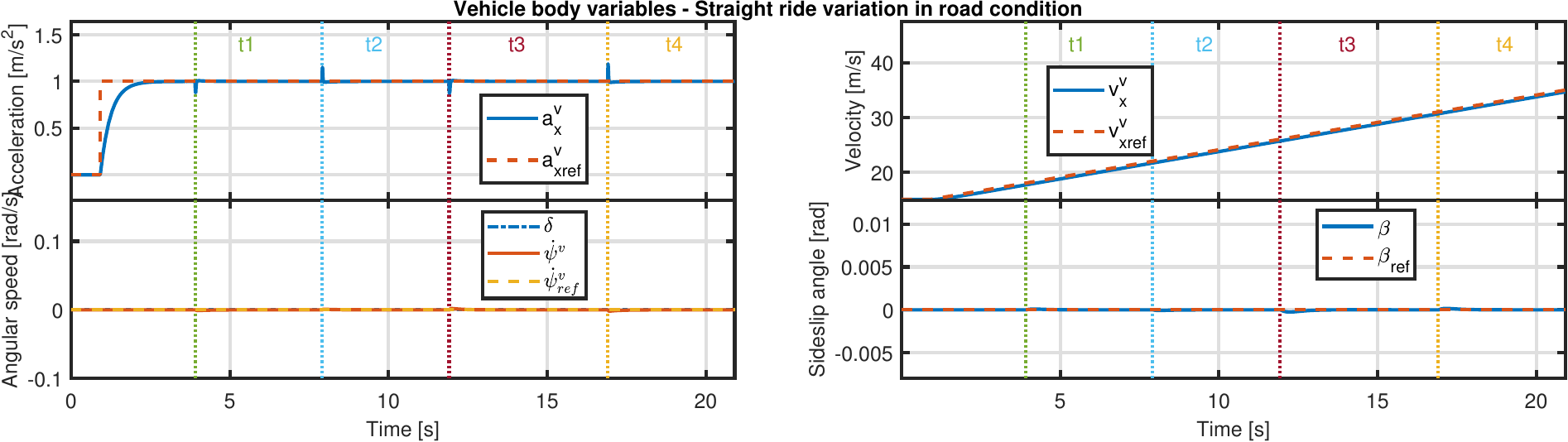"}
    \caption{Vehicle body-fixed variables ($v_{x}^{v}$, $\dot{\psi}^{v}$, $\beta$ and $a_{x}^{v}$) are shown here for the straight ride with $\epsilon_i$ variation (described in section \ref{sec:fz}).}
    \label{fig:fz_CG}
\end{figure*}
\begin{figure*}
    \centering
    \includegraphics[width=\linewidth]{"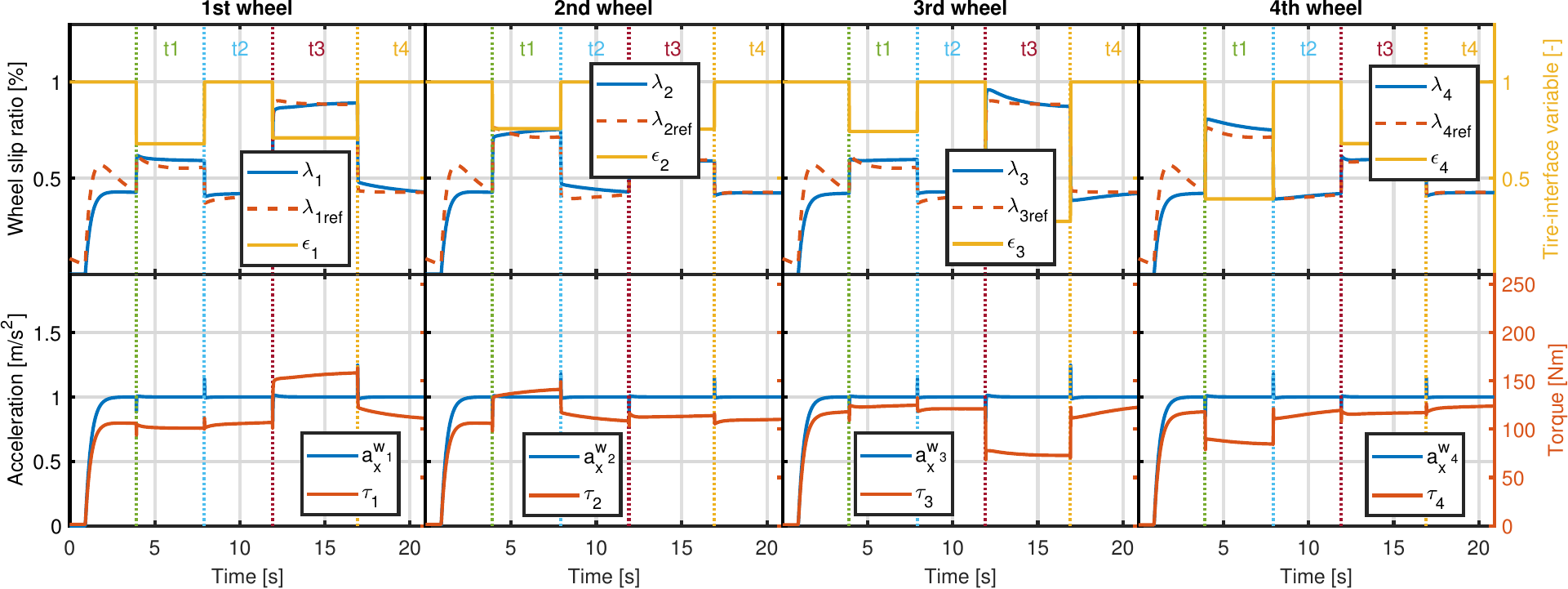"}
    \caption{Wheels variables ($a_{x}^{w_i}$, $\tau_{i}$, $\lambda_i$ and $\epsilon_i$ for each wheel) are shown here for the straight ride with $\epsilon_i$ variation (described in \ref{sec:fz}).}
    \label{fig:fz_w_i}
\end{figure*}
\begin{figure}
    \centering
    \includegraphics[width=.9\linewidth]{"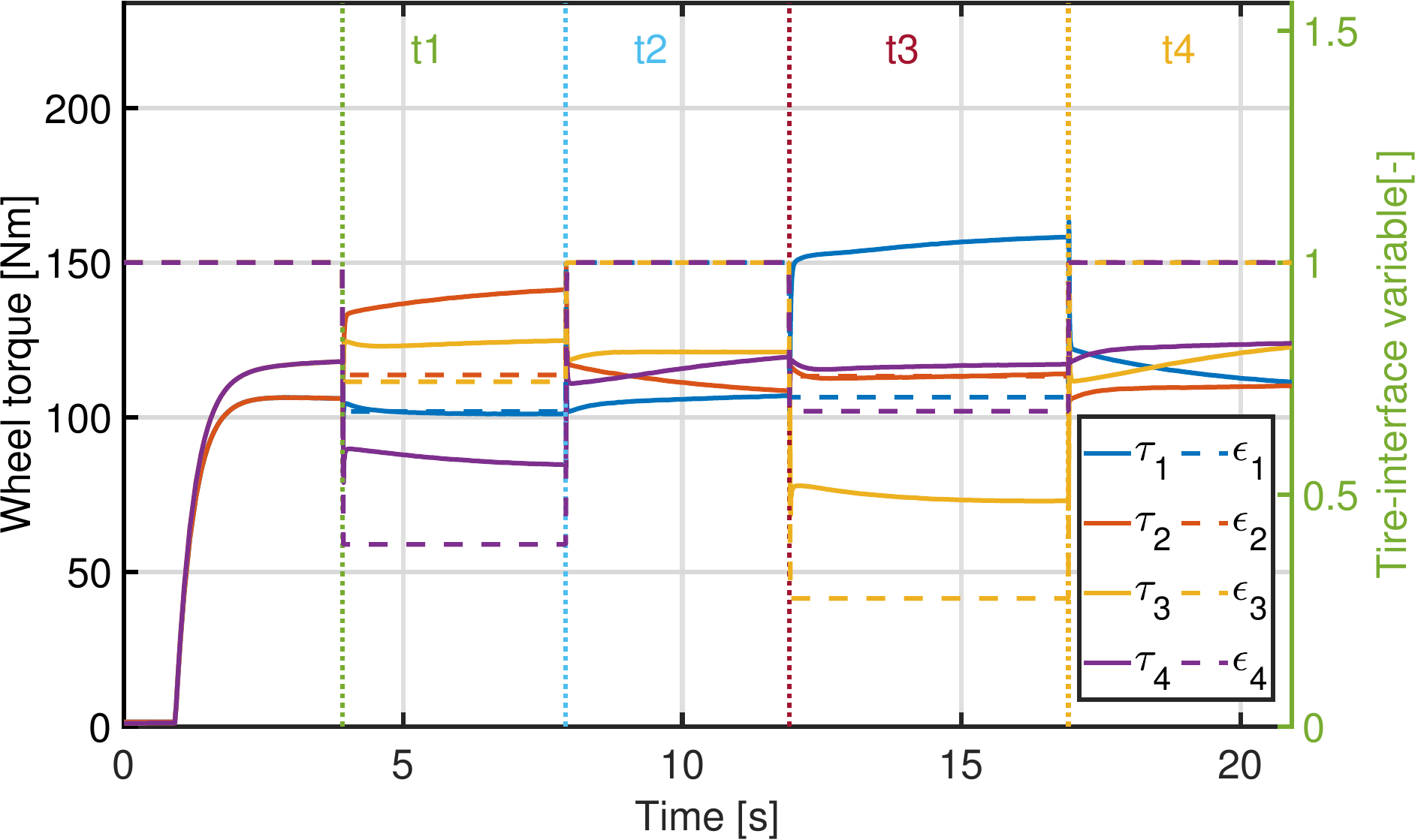"}
    \caption{Detail of wheel torques $\tau_i$ and tire-interface variables $\epsilon_i$ for each wheel are shown here for the Simulink model (see \ref{sec:model}) simulation experiment (see \ref{sec:fz}). The torque allocation based on $\mu_i$, $F_z{w_i}$, and $\epsilon_i$ can be seen.}
    \label{fig:fz_torques}
\end{figure}

\subsection{Description of MATLAB \& Simulink based experiment}
The experiment performed using the model introduced in \ref{sec:model} is described in this subsection. The effect of different surface types on the slip curve shape is approximated by slip curve scaling for the Matlab \& Simulink experiments. The slip curve shape is preserved and scaled down by an appropriate factor. 
The simplification does not bring a significant error as the slip curve is relatively flat around $\lambda_{max}$ and is assumed in many related works constant \cite{Maeda2012a, Zhai2016, Jia2018}. Furthermore, the presented control strategy is robust with respect to $\lambda_{\text{max}}$ position. The simplification is to be addressed in future work (different shapes of the slip curve can be seen in the figure \ref{fig:slip_curve}). Once having the $\lambda_{max}$ estimate, its integration into the controller proposed is straightforward.
The self-driven vehicle controller hierarchy (see \ref{sec:drive-by-human}) was used in this experiment.

The results are presented in a form consisting of graphs that are composed of:
\begin{itemize}
    \item \textbf{Vehicle body-fixed variables} - the response of vehicle fixed body represented by $v_{x}^{v}$, $\dot{\psi}^{v}$, $\beta$ and $a_{x}^{v}$ (acceleration of vehicle in $x$ direction (vehicle body-fixed CS)) variable.
    \item \textbf{Wheel variables} - the $\tau_{i}$, $\lambda_i$, $\epsilon_i$ and $a_{x}^{w_i}$ (acceleration of the wheel in $x$ direction - wheel CS) for each wheel are shown. 
\end{itemize}
\begin{figure*}
    \centering
    \includegraphics[width=\linewidth]{"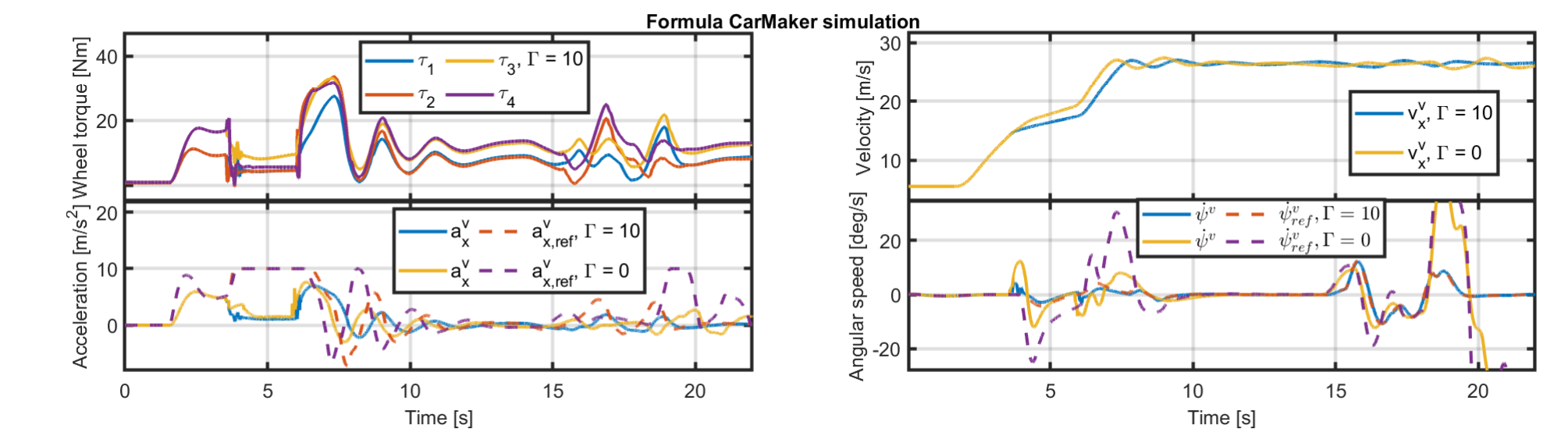"}
    \caption{Comparison of two lateral preference $\Gamma$ values for the CarMaker experiment described in \ref{sec:CarMakerExperiment}. The figures are from left to right in the first row: wheel torques and vehicle velocity. Next, in the second row: vehicle acceleration with its reference and vehicle yaw rate with its reference.}
    \label{fig:formula_GammaComp}
\end{figure*}
\begin{figure*}
    \centering
    \includegraphics[width=\linewidth]{"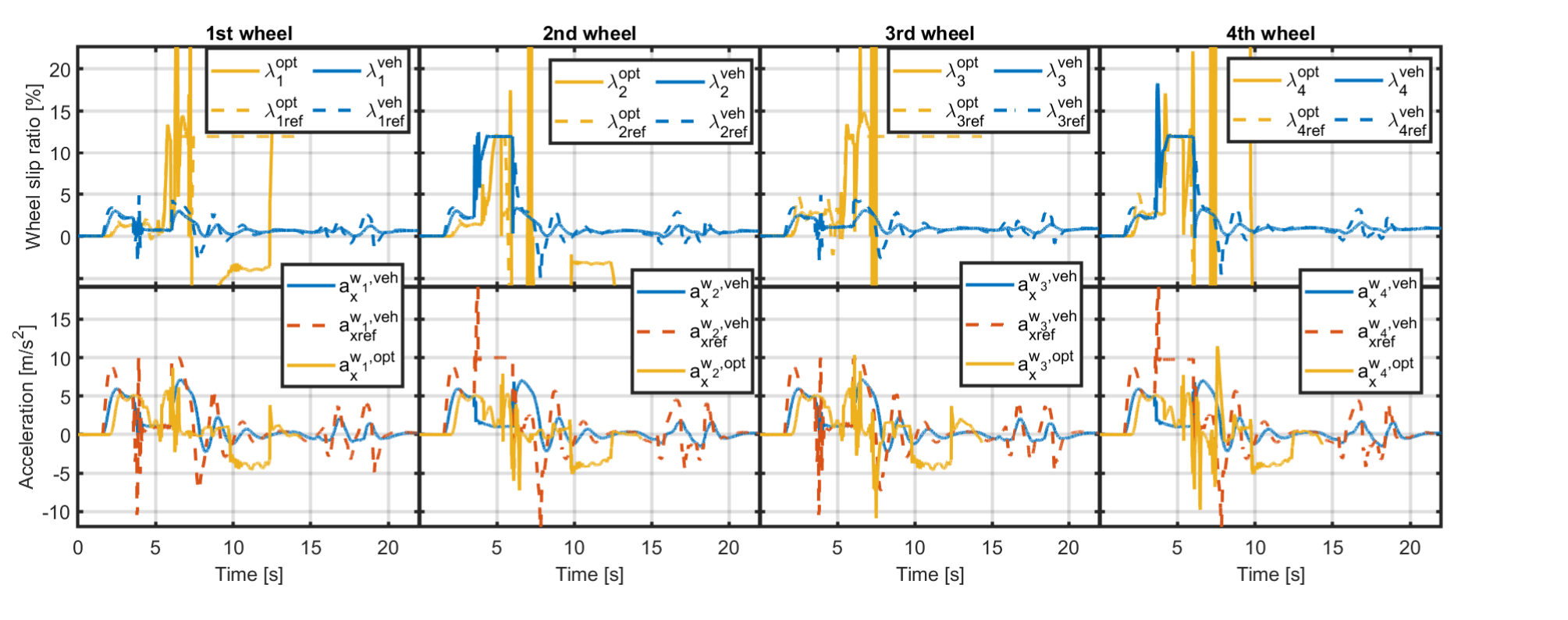"}
    \caption{Wheels acceleration $a_{x}^{w_i}$ and slip ratio $\lambda_i$ tracking comparison of the proposed allocation scheme (veh superscript is used) and the optimization based allocation scheme from \cite{Maeda2012a} (opt superscript is used). Details are described in \ref{sec:CarMakerExperiment}.}
    \label{fig:formula_RLSComp}
\end{figure*}
The following experiment was performed.
\subsubsection*{Straight ride - road condition variation}\label{sec:fz}
Artificial variation of the tire-to-road interface $\epsilon_i$ with constant vehicle acceleration while driving straight is simulated in the experiment.
The changes in $\epsilon_i$ are always different for each wheel. During this experiment:
\begin{itemize}
    \item  $\delta = 0\, rad$ is commanded through the whole experiment.
    \item $\epsilon_i$ is a piecewise constant and varies through the simulation as follows:
    \begin{itemize}
        \item $\epsilon_i = 1$ for times $t < t_1$, $t_2 < t < t_3$ and $t > t_4$
        \item $\epsilon_i$ is constant but different for each wheel otherwise ($\epsilon_i \neq 1$ and $\epsilon_i \neq \epsilon_j, i\neq j$)
    \end{itemize}
    \item Velocity reference is:
    \begin{itemize}
        \item $v_{x,\text{ref}}^{v}(t < 1) = 15\, m/s$
        \item $\frac{d}{dt}v_{x,\text{ref}}^{v}(t > 1) = 1\, m/s^2$
    \end{itemize}
\end{itemize} 
 The vehicle body-fixed variables can be seen in the \ref{fig:fz_CG}. Finally, the wheel variables are shown in \ref{fig:fz_w_i}. Torques $\tau_i$ and tire to road interface variables $\epsilon_i$ for each wheel are shown in figure \ref{fig:fz_torques}. The experiment outcomes are discussed later in the \ref{sec:benefits}.
\subsection{EFORCE formula CarMaker high fidelity verification experiment}\label{sec:CarMakerExperiment}
\begin{figure}
    \centering
    \includegraphics[width=0.95\linewidth]{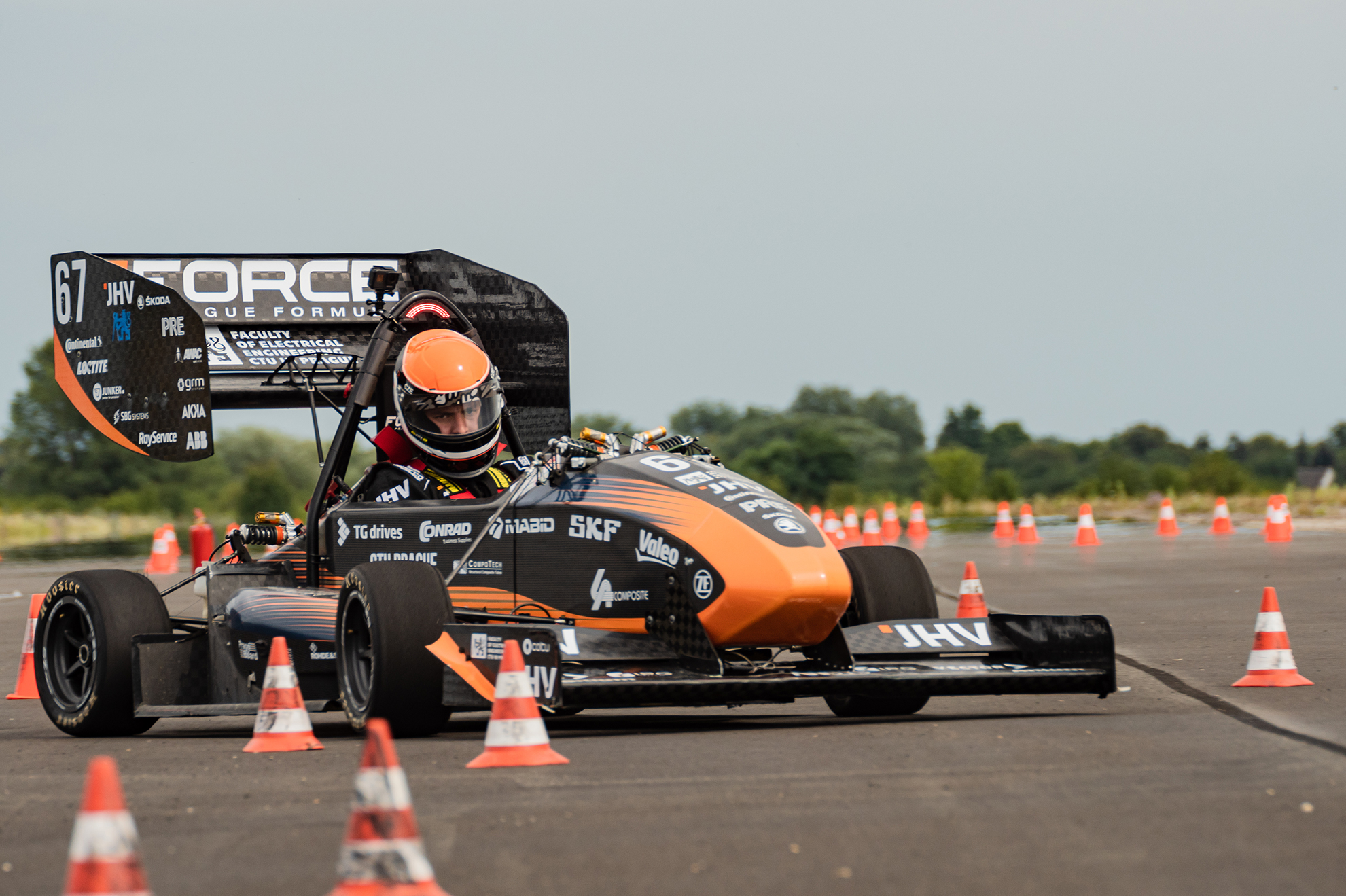}
    \caption{EFORCE student formula used for CarMaker model.}
    \label{fig:formula}
\end{figure}
The functionality and performance of the proposed hierarchical control system were validated in IPG CarMaker environment with high fidelity FEE CTU in Prague EFORCE student formula model containing all the dynamic phenomenas such as Tire models, Ackermann steering, load transfer effects, etc. The model is accessible at \cite{FormulaGit}.
The formula is shown in the \ref{fig:formula}. The calibrated CarMaker formula model with real parameters was used for validation. The drive-by-human controller hierarchy was used in the experiment (see \ref{sec:drive-by-human}). The IPG driver was used to control the pedals' positions translated to vehicle acceleration reference $a^{v}_{x~\text{ref}}$. The driver also controls the steering wheel angle $\delta$ translated to vehicle yaw rate reference $\dot{\psi}^{v}_{\text{ref}}$.  
The experiment consists of two parts -- acceleration and an ISO 3888-1 double lane change maneuver. The experiment starts with the vehicle maintaining a velocity of 20 km/h. Then, in time $t = 2s$ starts the vehicle acceleration. During the acceleration, the vehicle goes over a friction bump ($\mu$ = 0.15) with the left side wheels (approximately $t \in (4, 6)$). Finally, when reaching 95 km/h, an ISO double lane change maneuver is performed.

A comparison of three experiments with the setup described above is presented. $\Gamma = 0$ and $\Gamma = 10$ comparison is shown in \ref{fig:formula_GammaComp}. It can be seen that the $\Gamma = 0$ was unable to pass the double lane change at that high speed and the friction bump disturbance to the vehicle yaw rate tracking is higher than the $\Gamma = 10$ case. 
Comparison of $\Gamma = 10$ case and implementation of \cite{Maeda2012a} is shown in the \ref{fig:formula_RLSComp} for the wheel level variables. It can be seen that the \cite{Maeda2012a} was not even able to pass the friction bump. This is mainly due to the lack of traction limit handling compared to the proposed solution (see \ref{sec:tracLim}). Moreover, the solution of \cite{Maeda2012a} requires many other signals to be estimated or measured (like wheel normal force, etc.).
More simulation cases analysis is presented in the accompanying technical report \cite{FormulaGit}. Video is avialable at \cite{SDSYoutube2022}.
\subsection{Benefits of the proposed control system} \label{sec:benefits}
Benefits of the proposed control system are, among others
\subsubsection*{\textbf{Inherent control allocation}} The proposed transformation from the vehicle Center Point (CP) to any wheel pivot point is unambiguous  for vehicle state variables. As such, the set of wheel level reference signals is unique for any vehicle motion. The unambiguous properties of the transformation, in combination with feedback control of the vehicle/wheel state variables, resolve the control allocation problem. In contrast to the systems based on traction force control like \cite{Guo2018, Maeda2012a, Park2015, Li2020, Ge2011, Peng2019, Zhai2016} where the control allocation needs to be explicitly addressed. 
The experiment with variation of $\epsilon_i$ values (see \ref{sec:fz}) is presented in the \ref{fig:fz_w_i} and \ref{fig:fz_torques} (see times $t_1$ and $t_3$). Applied traction torque for each wheel is inherently adjusted by the control law to track the velocity reference signal and to reflect the road condition changes and, thus, the traction capabilities of the tire-to-road interface.
Finally, the traction torque inherent allocation is also shown in the CarMaker experiment (see \ref{fig:formula_GammaComp}).

\subsubsection*{\textbf{Robustness to center of gravity (CG) location and/or wheels steady-state normal loading}}
The velocity control is robust to wheel normal force/load variation. Furthermore, the proposed method inherently provides the wheel's normal force distribution ratio (with the assumption of uniform road friction properties) as a by-product. 
The effect of CG position and overall vehicle mass on traction torque allocation is presented in \ref{fig:fz_torques}. First, the vehicle accelerates with CG located rearwards of the vehicle CP. Values of rear axle torques $\tau_4$ and $\tau_3$ are adequately higher compared to front axle torques $\tau_2$ and $\tau_1$. The value of $\epsilon_i$ is changed at time stamps $t_1$ to $t_4$, representing a sudden shift of CG location and change of overall vehicle mass. The $\epsilon_i$ value scales the $\mu_i F_z^{w_i}D$ term (see eq.~\ref{eq:pacejka_long}), where the normal force $F_z^{w_i}$ depends on CG location and overall vehicle mass in steady-state drive.

\subsubsection*{\textbf{Robustness to dynamical load transfer}} Load of the wheel (normal force) is dynamically changing during the maneuvers. The vehicle velocity/acceleration control invariance to the CG location implies invariance to dynamic load transfer. Road grade and vehicle roll changes are interpreted as normal force variation, the same as dynamical load transfer. Moreover, the load transfer is modeled in the nonlinear simulation model (see \cite{TwinTrack}). 

\subsubsection*{\textbf{Robustness to tire-to-road interface variation}}
The vehicle velocity and mainly wheel angular speed feedback control provide robustness to handle road condition uncertainty. See \ref{sec:tireInterface} and \ref{sec:simulation} for more details. The $\epsilon_i$ value scales the $\mu_i F_z^{w_i}D$ term. From this point of view, $\epsilon_i$ models the road condition variations. Therefore, the figures mentioned in the pre-previous point are relevant for this point. It is especially clear in the \ref{fig:fz_torques}.

\subsubsection*{\textbf{Standard instrumentation}} The proposed system does not require any advanced instrumentation, both in the measurement and actuation sense, over the standard vehicles available today. The instrumentation level required is the same as that used for 4 e-motors vehicles with ABS and ESP systems. 

\subsubsection*{\textbf{Inherent wheel safety limits preservation}} The wheel traction is considered to be lost once the slip ratio is out of the $\lambda$ range where the derivative of slip curve is positive ($|\lambda| \geq \lambda_{max}$ in \ref{fig:slip_curve}, eq.~\ref{eq:pacejka_long}).  The $\lambda_{max}$ parameter value is imposed as a direct limitation in the presented wheel-level control law. It is assumed that the $\lambda_{max}$ parameter is to be known or estimated. There are papers concerning this challenging problem (see \cite{Seyedtabaii2019}). The $\lambda_{max}$ parameter estimation is not in the scope of the paper.
The proposed control system keeps the slip ratio $\lambda$ inside the prescribed boundaries (defined by $\lambda_{max}$ parameter). This functionality is ensured by the maximum $\lambda_{i,\text{ref}}$ value bounded in the $a_{x,\text{ref}}^{b_i}$ acceleration controller to the value $\lambda_{max}$.
\subsubsection*{\textbf{Unifying traction system design.}} Such a control strategy implements the functionality of all conventional wheel-level ADAS traction components like ABS, ESC, ASR, and others.
\subsubsection*{\textbf{System/vehicle level design}} The proposed traction control strategy is designed on the vehicle system level. This provides a higher/full integration level of traction functionality, higher performance, and robustness. The proposed algorithm's extremely low complexity maintains and often exceeds functionality performance. Finally, the low time required during the development and deployment reduces the overall cost of such systems.
\subsubsection*{\textbf{Yaw rate tracking and its preference to longitudinal dynamics tracking}} The proposed traction transformation introduces a designer tunable parameter influencing the yaw rate vs. longitudinal dynamics tracking preference.
This functionality is directly implied by the transformation \ref{eq:transform} and demonstrated in the EFORCE formula CarMaker experiment, see \ref{fig:formula_GammaComp} and \ref{fig:formula_RLSComp}. The slip ratio anti-symmetric action generating yaw rate $\dot{\psi}^{v}$ is clearly visible between $15^{th}$ and $20^{th}$ second in the \ref{fig:formula_GammaComp}. Thus, the Torque Vectoring functionality is inherently implemented by the proposed control architecture. 
The yaw rate to longitudinal dynamics tracking preference was tested in the formula double lane change maneuver. The difference between $\Gamma = 10$ and $\Gamma = 0$ is clearly visible in figure \ref{fig:formula_GammaComp} (yaw rate tracking).


\section{Conclusion}
This paper presented a vehicle motion control system with many benefits compared to state-of-the-art systems. References at the vehicle center point are transformed into individual wheel pivot points references uniquely. A hierarchical control system unit that tracks these setpoints is proposed here for each wheel (see \ref{sec:controller}). The safety and performance of the whole system were shown in the Simulink and CarMaker simulations with EFORCE formula (see \ref{sec:simulation}). The simulations were executed using the nonlinear twin-track model (see \ref{sec:model}) for Simulink experiments and with the CarMaker validation formula model in the CarMaker environment. The benefits of the designed model are listed in the \ref{sec:benefits}. The proposed control system was also compared against the state-of-the-art approach \cite{Maeda2012a}.

\appendices


\ifCLASSOPTIONcaptionsoff
  \newpage
\fi

\bibliographystyle{myieeetran}
\bibliography{vosahlik_ref,ostatni}

\begin{IEEEbiography}[{\includegraphics[width=1in,height=1.25in,clip,keepaspectratio]{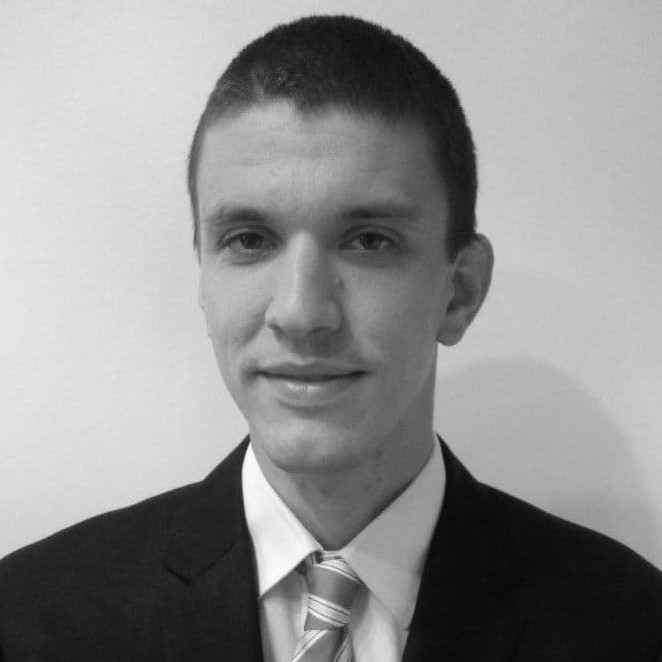}}]{David Vošahlík}
	received the master degree in cybernetics and robotics from the Faculty of Electrical Engineering (FEE), Czech Technical University (CTU) in Prague, in 2020. He is currently working towards the Ph.D. degree in advanced vehicle dynamics control at the Department of Control Engineering, FEE, CTU in Prague. His research interests include vehicle dynamics control with special focus on safety and economy.
\end{IEEEbiography}

\begin{IEEEbiography}[{\includegraphics[width=1in,height=1.25in,clip,keepaspectratio]{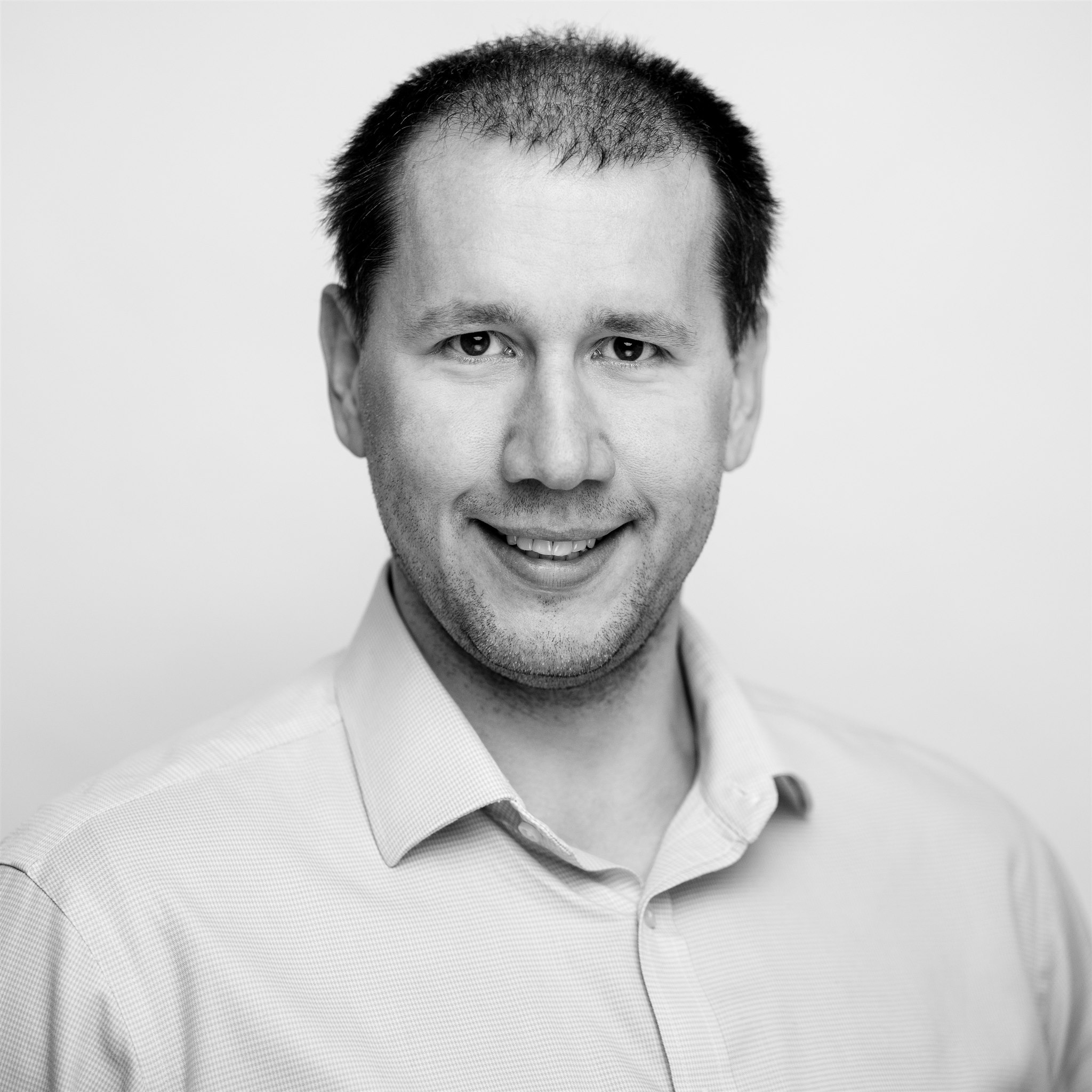}}]{Tomáš Haniš}
	did his Ph.D. thesis on the topic of advance control system design for flexible structure aircraft at CTU in Prague with an internship at EADS (Airbus) and Space Dynamics Laboratory in Utah, followed by 2 years Postdoc on the topic of Optimization based advance control system design for Wind-turbines with flexible structures at DTU.
	Tomas has industrial experience in the automotive sector, as he was holding the post of Team leader and control system design competence owner at Porsche Engineering Group, as well as the aerospace sector as a control system design engineer at Rolls-Royce Aerospace.
	Tomas is currently in charge of Smart Driving Solutions research center at Department of Control Engineering, Faculty of Electrical Engineering, Czech Technical University in Prague.
\end{IEEEbiography}


\end{document}